\documentclass[lettersize,journal,twoside]{IEEEtran}
\usepackage{amsmath,amsfonts}
\usepackage[ruled,linesnumbered]{algorithm2e}
\SetAlFnt{\small}
\SetKwInput{KwInput}{INPUT}
\SetKwInput{KwOutput}{OUTPUT}
\usepackage{amsmath}
\usepackage{colortbl}
\usepackage{array}
\usepackage{mathtools}
\usepackage{makecell}
\usepackage{textcomp}
\usepackage{stfloats}
\usepackage{minted}
\usepackage{url}
\usepackage{verbatim}
\usepackage{graphicx}
\usepackage{cite}
\usepackage{xcolor}
\usepackage{xspace}
\usepackage{tcolorbox}
\usepackage{lipsum} 
\usepackage{threeparttable}
\usepackage[
    font=normalsize,
    labelfont=sf,
    textfont=sf
]{caption}
\usepackage{subcaption}
\usepackage{booktabs} 
\usepackage{multirow}
\usepackage{pgfplots}
\pgfplotsset{compat=1.18}
\usepackage{hyperref}
\usepackage[hyphenbreaks]{breakurl}
\definecolor{mygreen}{RGB}{0,100,0}
\definecolor{rqColorA}{HTML}{38b775}
\definecolor{rqColorB}{HTML}{f8e620}
\SetCommentSty{textit}

\SetCommentSty{mycommentstyle}

\newcommand{\ignore}[1]{}
\newcommand{\tool}{{\scshape{SiblingRepair}}\xspace}
\newcommand{\hercules}{{\scshape{Hercules}}\xspace}

\begin{document}

\title{\tool: Sibling-Based Multi-Hunk Repair with Large Language Models}

\author{Xinyu Liu, Jiayu Ren, Yusen Wang, Qi Xin, Xiaoyuan Xie, and Jifeng Xuan
\thanks{All authors are with the School of Computer Science, Wuhan University, Wuhan 430072, China
(e-mail: xinyuliu4@whu.edu.cn; renjiayu@whu.edu.cn; yusenwang@whu.edu.cn;
qxin@whu.edu.cn; xxie@whu.edu.cn; jxuan@whu.edu.cn).
Corresponding author: Qi Xin.}
}

\maketitle

\begin{abstract}
Developers often make similar mistakes across multiple code locations implementing related functionalities.
These buggy locations, referred to as siblings, share similar issues and require similar fixes.
It is important for automated program repair (APR) to identify such siblings and
apply consistent patches for their repair.
The state-of-the-art \hercules is designed for this purpose
but it is limited by strong assumptions about sibling locations and commit-history availability,
rigid AST-based sibling matching that may miss valid siblings,
and template-based patch generation with low flexibility and effectiveness.

To address these limitations, we propose \tool,
a new LLM-based multi-hunk APR technique that specializes in sibling repair.
Starting from a suspicious location identified by spectrum-based fault localization,
\tool searches for semantically related sibling candidates
using token- and embedding-based code matching without restricting sibling discovery
to failing-test coverage or commit history.
It then leverages an LLM to identify true failure-relevant siblings
and generate consistent patches via two strategies: 
simultaneous repair, which jointly identifies and repairs siblings,
and iterative repair, which proceeds in a progressive manner by 
examining each sibling candidate and evaluating the generated patches
based on test outcomes and execution feedback.
\tool also preserves promising patches identified in previous suspicious location repair 
and carries them forward; these patches across multiple iterations 
can be combined to form a general multi-hunk patch.

We implemented a prototype of \tool and evaluated it on Defects4J and GHRB.
Compared with state-of-the-art multi-hunk APR techniques including \hercules and ITER,
\tool achieves better repair. 
For example, under spectrum-based fault localization,
it correctly repaired 
23 sibling bugs and 76 multi-hunk bugs in Defects4J.
By comparison, \hercules repaired only 4 sibling and 9 multi-hunk bugs,
and ITER repaired only 24 multi-hunk bugs.
We also show that \tool is efficient at generating correct patches,
and that its candidate sibling detection and LLM-based repair components are effective,
and that its performance with different LLMs on GHRB is comparable,
which implies limited data leak impact for LLM-based repair.
Overall, these results demonstrate \tool's substantial advancement
for sibling-based and general multi-hunk repair.

\end{abstract}


\markboth{}%
{Liu \MakeLowercase{\textit{et al.}}: \tool: Sibling-Based Multi-Hunk Repair with Large Language Models}


\section{Introduction}

Automated program repair (APR)~\cite{goues2019automated,zhang2024systematic} is gaining momentum nowadays.
It holds the promise of automatically fixing bugs to save developers' time and effort for debugging.
Developers often make similar mistakes~\cite{islam2016bug,saha2019harnessing} at different locations.
As a result, a significant portion of real fixes handle related issues.
This is evidenced by a study~\cite{xin2024detecting} that shows
about 11\% of multi-hunk fixes addressing multiple locations deal with related issues.
Our investigation of the developer patches (ground-truth)
of the Defects4J bugs~\cite{just2014defects4j} also shows that 
14.2\% of the multi-hunk fixes are similar, according to GumTree's code matching algorithm~\cite{falleri14fine,falleri24fine}.
Thus, it is important for APR to generate similar fixes to
repair the related issues across multiple locations.
Following previous work~\cite{saha2019harnessing}, we refer to code locations exhibiting related issues 
and requiring similar fixes as \textit{siblings}, 
the corresponding bugs as \textit{sibling bugs}, and the repair
addressing such bugs as \textit{sibling repair}.

\hercules~\cite{saha2019harnessing} is a state-of-the-art APR technique designed for sibling repair.
Given a buggy program and a test suite comprising passing tests
and at least one failing test exposing the failure,
\hercules assumes that the failure is due to a sibling bug,
and for its repair, \hercules starts with spectrum-based fault localization
to identify a list of suspicious locations.
For each location $l$, \hercules leverages test coverage, commit history,
and an AST-based code matching algorithm to identify its siblings. 
To repair the siblings alongside $l$, \hercules uses template-based patch generation.

Although promising with its stated goal, \hercules suffers from three key weaknesses:
(1) strong assumptions on where the siblings exist and the usefulness of commit-history;
(2) rigid AST-based matching algorithm for sibling matching; and
(3) template-based patch generation, which has limited repair ability 
and has no tolerance of any sibling identification errors. 

Specifically, for (1), \hercules assumes that the sibling locations
are covered in the execution of failing tests, and when missing,
can be identified from the commit history, since siblings can co-occur in historical revisions.
Unfortunately, failing tests may not be sufficient to
capture all similar mistakes but often one or a few of them.
Besides, their executions can terminate prematurely due to early exceptions or assertion failures.
As a result, failing test executions do not necessarily cover all the siblings,
as also acknowledged by \hercules's authors~\cite{saha2019harnessing}.
Our own investigation of the test coverage also shows that
for as high as 61\% of the Defects4J sibling bugs, failing tests do not cover all siblings.
Although the commit history might help with sibling identification,
its actual usefulness is limited. As evidence, we found that
for only 17\% of sibling bugs where the test coverage is insufficient,
the commit history can complement this by containing the co-occurrence records 
for all siblings.

For (2), \hercules uses an AST-based code matching algorithm
that relies on code structure comparison, in particular,
the compatibility of node kinds (following super- or sub-class relationships) and
types (satisfying type casting criteria) for sibling identification.
However, siblings that share a related issue may not be
kind- or type-compatible, as the core buggy part may be wrapped in
different types of expressions or statements that are not
compatible under \hercules's criterion. As such, \hercules may incorrectly exclude siblings for repair.

For (3), once the siblings are identified,
\hercules uses template-based patch generation to repair them.
However, it is well known that a limited set of templates are insufficient 
to handle diverse forms of buggy code~\cite{meng2023template,huang2025template}.
Besides, \hercules's patch generation has no tolerance of sibling identification errors --- 
it only attempts to repair all identified siblings together,
so any mis-identification can lead to an ineffective patch.

To overcome these weaknesses, we propose \tool, 
a new LLM-based APR technique that specializes in sibling repair
while also supporting general multi-hunk repair.
Given a buggy program and the test suite, \tool performs spectrum-based fault localization
to identify a list of suspicious locations and, for each,
identifies siblings and generates patches for repair.
Unlike \hercules, \tool does not assume the siblings are covered by failing tests only,
nor does it rely on commit history for sibling identification.
Instead, \tool allows siblings to be anywhere in the code
exercised by any test (not just the failing tests).
To identify them, \tool collects statements exercised by the test suite,
extracts their surrounding code contexts, and performs code comparison
based on semantic embeddings to identify locations related to 
the suspicious location as candidate siblings.
Since semantic embedding comparison over all exercised statements is expensive, 
\tool adopts a more feasible method: it first performs token-based matching
using TF-IDF similarity~\cite{leskovec2020mining} to exclude
low-similarity locations that are unlikely to be siblings, 
and then applies semantic comparison only to the remaining candidates. 
\tool ignores potential siblings not exercised by any test, 
since repairs of such siblings cannot be validated by existing tests
and may introduce unsafe changes.

With the candidate siblings found, \tool next determines failure-relevant siblings
and generates patches with the help of an LLM.
To this end, \tool adopts two complementary strategies: simultaneous repair and iterative repair.
In simultaneous repair, the goal is to jointly identify and repair the siblings.
To support this process, \tool constructs a prompt that includes the suspicious location 
and its enclosing method, the failing test case,
the failure information (including error message and stack traces),
a filtered set of the candidate siblings (to fit the prompt window),
relevant fix ingredients, and the test feedback.
This prompt provides the LLM with the bug context,
failure evidence, and sibling candidates for identification,
while also guiding patch generation using the provided fix ingredients. 
In addition, test feedback is used to help the LLM refine the generated patches.

Although simultaneous repair enables a joint fix for all failure-relevant locations,
it may fail due to sibling filtering errors or LLM's hallucination issues,
which cause inaccurate sibling identification and incorrect patch generation for the shared issues.
To mitigate these limitations, \tool also performs iterative repair, in which
it examines candidate siblings one at a time. For each candidate sibling,
it prompts the LLM to repair only that location and then evaluates whether
the generated patch leads to any progress, as indicated by reduction in failing tests
or an extended execution (e.g., progressing beyond a previous failure point).
When such progress is observed, \tool treats the sibling as failure-relevant and 
reuses the resulting patch in subsequent prompts to guide repairs at other locations.

Although \tool specializes in sibling repair, it also supports 
general single-hunk and multi-hunk repair.
Given a suspicious location and its candidate siblings,
the LLM may determine that only the suspicious location requires a modification,
yielding single-hunk patches. For multi-hunk repair,
\tool retains promising patches generated at earlier suspicious locations and
carries them over to subsequent locations. When combined, these promising patches 
can form a general multi-hunk patch.

To assess the effectiveness of \tool, we implemented a prototype and evaluated it on
two bug benchmarks, Defects4J~\cite{just2014defects4j} and GHRB~\cite{lee2024github}.
We compared \tool with four state-of-the-art multi-hunk APR techniques 
including \hercules~\cite{saha2019harnessing} and ITER~\cite{ye2024iter}.
Our results demonstrate \tool's superiority in repairing sibling bugs:
using the DeepSeek LLM (v3.2), it successfully repaired 23 
Defects4J sibling bugs under SBFL, whereas \hercules repaired only 4.
Despite specializing in sibling repair, 
\tool repaired 76 multi-hunk bugs and 100 single-hunk bugs
in Defects4J with correct patches, substantially outperforming the baseline techniques,
with the best repairing only 78 bugs (multi-hunk and single-hunk).
Under single perfect fault localization,
where a real buggy location is known and 
placed at the top of the suspicious list, 
\tool can repair many more bugs, or 278 in total.
Our results also show that \tool is efficient at generating plausible patches 
and that its candidate sibling detection and
its complementary repair components are effective.
Moreover, our evaluation of \tool on the GHRB sibling bugs
shows that it achieves comparable results with different LLMs,
GPT-3.5-Turbo, whose training predates the dataset,
and DeepSeek v3.2, a more recent model. This result implies
a limited data leakage impact on \tool's effectiveness.
Taken together, these results demonstrate \tool as a strong
sibling-based APR technique for multi-hunk repair.

In summary, our paper makes the following contributions.

\begin{itemize}

\item An LLM-based multi-hunk APR technique \tool that
identifies and repairs sibling bugs using a combination of 
token- and embedding-based matching with simultaneous and iterative repair strategies, 
while also supporting general multi-hunk repair.

\item A comprehensive evaluation showing that \tool repairs 
more sibling bugs than the state-of-the-art technique \hercules and
outperforms other techniques in general multi-hunk repair.

\end{itemize}

\section{Motivation and Overview}

In this section, we use a motivating example to illustrate the limitations of \hercules, 
and then provide an overview of \tool,
showing how it addresses these limitations.

\subsection{Motivating Example}
\label{subsec:motivating example}

\begin{figure*}[t]
    \centering
    \includegraphics[width=\linewidth]{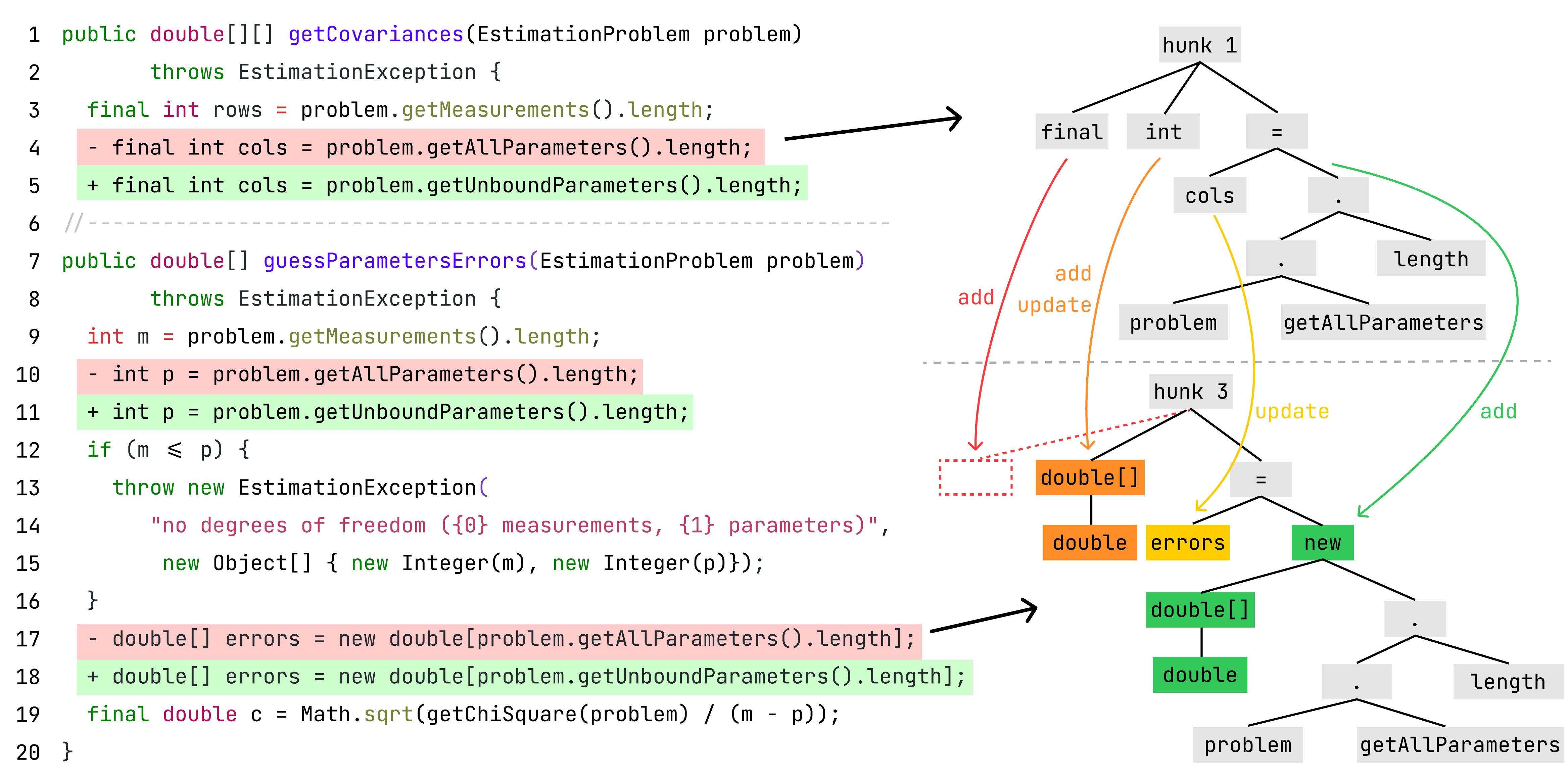}
    \caption{The Math\_100 bug and its correct patch (on the left),
    along with \hercules's AST-based code matching and edit generation (on the right).}
    \label{fig:math100}
\end{figure*}

We use the Math\_100 bug from Defects4J~\cite{just2014defects4j} as a motivating example.
The failure manifested as \textit{ArrayIndexOutOfBoundsException}
is caused by using the full parameter vector in the computation
that expects only the active (unbound) parameter subset, leading to a dimension mismatch 
and out-of-bounds access.
To fix the bug, the developer patch consistently replaces \textit{getAllParameters()} 
with \textit{getUnboundParameters()} at three locations, as shown in Fig.~\ref{fig:math100}.

This example illustrates a class of sibling bugs that require similar fixes
at multiple locations with contexts not highly similar.
\hercules fails to repair this bug due to three key limitations:
(1) strong assumptions about the sibling locations
and the usefulness of commit history; (2) rigid AST-based code matching method
for sibling identification; and (3) the limited flexibility and repair ability
of its template-based patch generation.

First, \hercules searches for siblings within code exercised by failing tests.
However, in this example, the failing test alone is insufficient to
cover all siblings, leading to false negatives.
Specifically, the failing test execution terminates prematurely
due to the \textit{ArrayIndexOutOfBoundsException} triggered by
a call to \textit{getCovariance}, and therefore does not complete.
As a result, only the sibling at line~4 is covered, 
whereas the other two at line~10 and line~17 are not.
While \hercules also seeks co-occurring relationships from commit history for sibling identification,
the history provides no evidence of co-occurrence in this case,
since the three locations were never modified together.
As a result, \hercules incorrectly excludes the siblings at lines~10 and~17,
leading to a repair failure.

Second, even if \hercules were to include lines~10 and~17 as sibling candidates,
its AST-based code matching algorithm would still fail to 
identify all three locations as failure-relevant siblings to repair.
This is because the algorithm relies heavily on structural similarity, comparing the compatibility of
node kinds (based on super- or sub-class relationships) and types (according to type casting criteria).
However, siblings may differ substantially in statement structure 
even when their core buggy expressions are similar.
In this example, the statement at line~17 differs structurally
from those at lines~4 and~10, causing the algorithm to exclude it as a sibling.
Specifically, the statement at line~4 is a scalar variable declaration,
whereas the one at line~17 is an array declaration. Moreover,
the right-hand side of line~4 is a field access, whereas that of line~17
is an array initialization. These structural differences lead to
an edit distance of 7 in statement comparison 
under \hercules's employed tree distance algorithm~\cite{zhang1989simple}, 
ultimately causing line~17 to be excluded.

Third, \hercules uses template-based patch generation, inheriting
11 templates from existing work~\cite{kim2013automatic,saha2017elixir,xiong2017precise}, 
such as \textit{Insert Cast/Null Pointer/Range Checkers},
\textit{Mutate Operators/Return Statements}, and \textit{Statement Movement}. 
However, a limited set of templates are not effective to deal with diverse forms 
of buggy code. Besides, the patch generation is inflexible, in that
it has no tolerance of sibling identification errors:
\hercules only attempts to apply templates to all the identified siblings
for patch generation, and any identification error would result in repair failures.

\subsection{Overview of \tool}
\label{sec:sys_overview}

\tool addresses these limitations with token- and embedding-based
candidate sibling identification and a combination of simultaneous
and iterative strategies for sibling matching and repair with the help of an LLM.

\begin{figure*}[t]
    \centering
    \includegraphics[width=\linewidth]{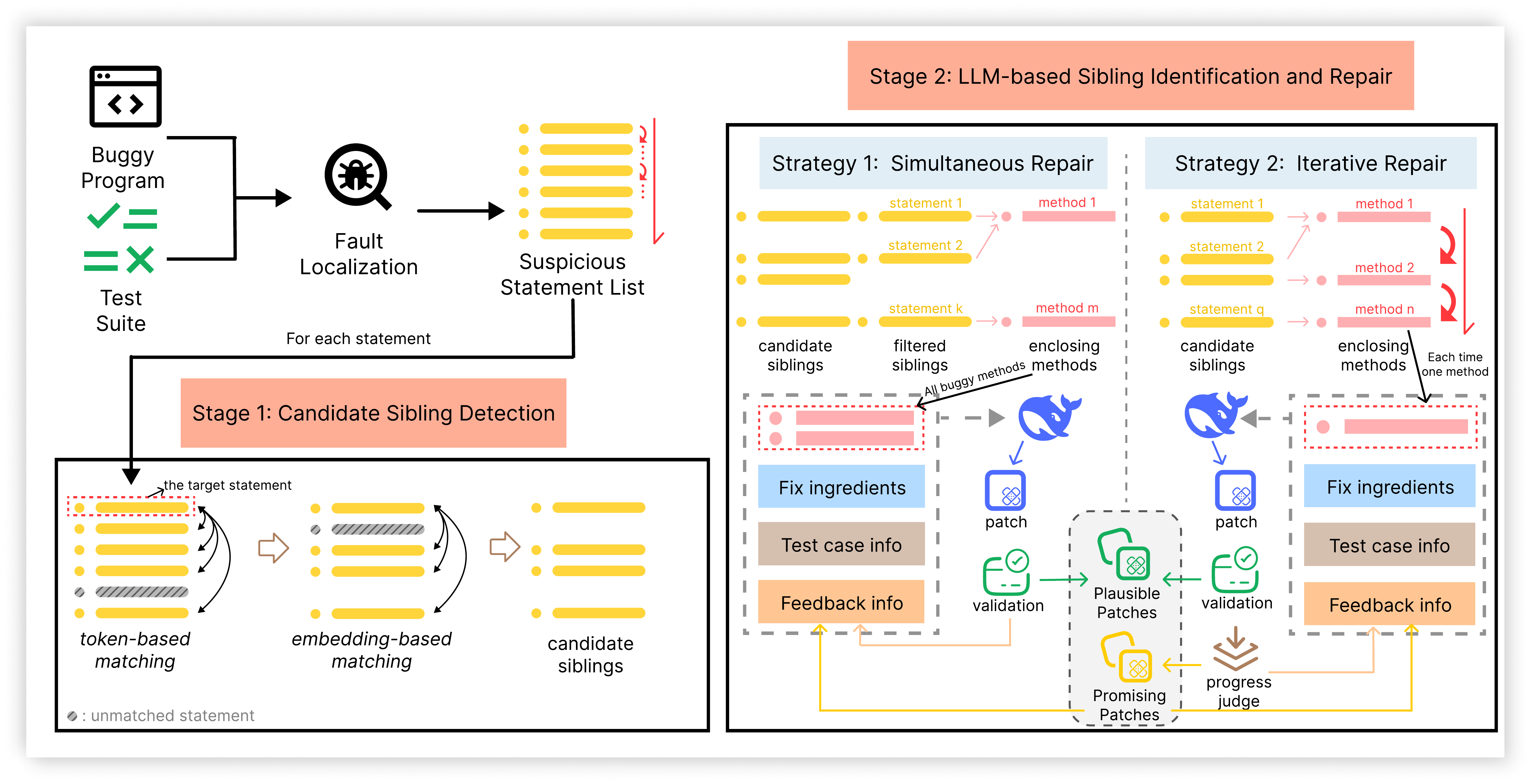}
    \caption{The overview of \tool. 
    }
    \label{fig:overview}
\end{figure*}

The workflow of \tool is illustrated in Fig.~\ref{fig:overview}.
Given a buggy program and the test suite, \tool first performs
spectrum-based fault localization (SBFL) to obtain a ranked list of suspicious statements.
For each, \tool operates in two stages for sibling repair: 
candidate sibling detection and LLM-based sibling identification and repair. 
In the first stage, \tool identifies candidate siblings of the suspicious statement. 
Then, in the second stage, \tool uses two strategies:
simultaneous repair and iterative repair to identify failure-relevant siblings
and repair them. 
We next walk through the two stages with the example to explain how \tool works.

Starting from the suspicious statement at line~4 (ranked 9th by fault localization),
\tool first searches for semantically related sibling candidates.
Unlike \hercules, it does not restrict the search to code exercised 
only by failing tests, nor does it rely on commit history.
Instead, it considers all code locations exercised by any test,
yielding a pool of 639 potential sibling locations in this example.
Rather than relying on rigid structural matching, \tool aims to identify semantically related code.
To this end, it first extracts the local context of line~4---which in this case
contains line~4 and its previous line using \hercules's method~\cite{saha2019harnessing}---and
applies TF-IDF similarity to filter out locations unlikely to be siblings.
It then performs embedding-based semantic comparison only on the remaining candidates,
thereby reducing the overhead of semantic embedding vector generation.
Through this two-stage filtering and semantic matching process, 
\tool identifies 73 candidate statements,
including the true, failure-relevant sibling locations at lines~10 and~17.

Once the candidate siblings are identified, \tool next performs simultaneous repair,
leveraging an LLM to jointly identify the failure-relevant siblings and repair them.
Unlike \hercules's template-based patch generation,
which applies predefined templates to all candidates and is highly sensitive to identification errors,
the LLM can flexibly select locations to repair
based on its reasoning about the root cause. Besides, its patch construction ability
is more powerful than a template-based method.
In this step, \tool prompts the LLM with the suspicious location and its enclosing method,
the failing test, the failure information, the candidate siblings,
and relevant fix ingredients (fields and methods defined in related classes)
to identify the failure-relevant siblings and generates method-level patches.
It also uses previous test failures and generated patches as feedback
to help the LLM refine its results.
To ensure that the candidate siblings fit within the prompt window,
\tool applies an additional filtering step based on Jaccard similarity~\cite{leskovec2020mining},
which in this example reduces the candidate set from 73 to 24
(across 13 methods). Through simultaneous repair, \tool successfully
identifies the two methods containing the three statements in red 
as the target repair locations and constructs the correct patch shown in Fig.~\ref{fig:overview}.
In particular, the retrieved fix ingredient \textit{getUnboundParameters()}
from the relevant class of the parameter \textit{problem} helps with patch construction.

Although simultaneous repair suffices for this example,
it may fail in general due to either sibling filtering errors, 
which can lead to incomplete repairs, or LLM hallucination issues,
which can result in incorrect sibling identification and patch generation.
We observed such failures in the JacksonDatabind\_32 and Math\_46 bugs from Defects4J.
For JacksonDatabind\_32, \tool mistakenly filters out a failure-relevant sibling
that is not highly similar to the suspicious location but requires
a similar fix: the insertion of \texttt{case JsonTokenId.ID\_END\_OBJECT:}.
For Math\_46, the LLM incorrectly identifies \texttt{return NaN;}
in the \texttt{atan} method as a failure-relevant sibling, likely because
it shares similar NaN-handling semantics with the real siblings
and \texttt{atan} invokes the method where the real siblings exist.

To address these limitations, \tool also performs iterative repair,
in which it examines candidate siblings one at a time,
generates method-level patches only for each candidate individually,
and evaluates whether any generated patch is promising.
This evaluation is based on test feedback and stack traces,
checking whether a patch reduces failing tests
or possibly extends execution beyond a previous termination point
(e.g., an exception-throwing location). 
Candidate locations for which no promising patch is found 
are considered not failure-relevant and are skipped.
Promising patches generated from earlier candidate locations
are then reused to guide the repair of subsequence locations.
For both JacksonDatabind\_32 and Math\_46,
\tool successfully repaired the bugs through iterative repair
by reusing promising patches retained to guide subsequent repair
while correctly skipping irrelevant locations.
Moreover, these retained patches not only support sibling repair 
for the current suspicious location but are also carried forward 
to subsequent suspicious locations, enabling general multi-hunk repair.

\section{The \tool Approach}
Algorithm~\ref{alg:siblingsrepair_main} outlines the workflow of \tool.
The algorithm begins with spectrum-based fault localization (SBFL)
using the common Ochiai metric~\cite{abreu2007accuracy} for suspiciousness quantification (line~4).
For each suspicious location and its context (line~6),
it searches for candidate siblings using token- and embedding-based matching (lines~7--8).
It then performs simultaneous repair (line~9) and iterative repair (line~13) 
with the LLM to identify failure-relevant siblings and generate patches.
Promising patches are retained in $P_{pro}$ and carried forward
across the iterations of suspicious location repair.
We next elaborate on the key steps: candidate sibling detection,
simultaneous repair, and iterative repair.

\SetKwProg{Fn}{}{}{end}
\SetKw{KwInWord}{in}
\begin{algorithm}
    \caption{The \tool Algorithm}
    \label{alg:siblingsrepair_main}
    \KwInput{initial buggy program $b$, test suite $T$, fault localization $F$, max candidate siblings number $k$, embeding similarity threshold $\theta$, jaccard similarity threshold $\alpha$, repair attempts $t$, max fix ingredients of each line $n$}
    \KwOutput{plausible patches $P_{pl}$}
    \Fn{\tool $(b, k, \theta, \alpha, t, n, T, F)$}{
        $P_{pro} \gets \emptyset;$ \tcp{initial promising patch}
        $P_{pl} \gets \emptyset;$ \tcp{initial plausible patch}
        $L \gets F(b, T);$ \tcp{suspicious statements list}
        $L_{\text{ctx}} \gets \textsc{ContextExtend}(L);$
        
        \For{$l_{\text{ctx}} \;\mathrm{in}\; L_{\text{ctx}}$}{
            $L_{\text{ctx}\_\text{rel}\_\text{init}} \gets \textsc{TokenBasedMatching}(l_{\text{ctx}}, L_{\text{ctx}}, k)$; \tcp{statement with relevant context}
            ${L_{\text{ctx}\_\text{rel}}} \gets \textsc{EmbeddingBasedMatching}(l_{\text{ctx}}, L_{\text{ctx}\_\text{rel}\_\text{init}}, \theta)$\; 
            $\textsc{SimRepair}({L_{\text{ctx}\_\text{rel}}}, b, t, \alpha, n, P_{pro}, P_{pl}, T)$\;
            \If{$P_{pl} \neq \emptyset$}{
                \textbf{break}\;
            }
            $\textsc{IterRepair}({L_{\text{ctx}\_\text{rel}}}, b, t, n, P_{pro}, P_{pl}, T)$\;
            \If{$P_{pl} \neq \emptyset$}{
                \textbf{break}\;
            }
        }
        \Return{$P_{pl}$}\;
    }
\end{algorithm}

\subsection{Candidate Sibling Detection}
With fault localization, \tool obtains a ranked list of suspicious statements. 
For each suspicious statement, it identifies candidate siblings --- 
locations that potentially share similar underlying issues (e.g., related API misuses) and
require similar fixes. To achieve this, \tool adopts a three-step procedure:
(1) context extraction, (2) token-based matching, and (3) embedding-based matching. 

\subsubsection{Context Extraction}
\tool adopts \hercules's method~\cite{saha2019harnessing},
which is based on reaching-definition analysis, to extract context
for both the suspicious statement and each potential sibling (exercised by tests).
Specifically, it first collects all variables used in the target statement
representing the suspicious location and then includes their most recent assignments or definitions
as the relevant context. If no such context is found,
the statement immediately preceding the target statement is used instead.
Note that the context extracted includes the target statement itself.

\subsubsection{Token-based Matching} 
Directly applying embedding-based comparison across all potential locations
is computationally expensive, given the costly interaction with 
the online model TEXT-EMBEDDING-ADA-002~\cite{greene2022embedding} for embedding generation, 
even in batch mode (approximately 2-3 hours for 10,000 locations).
To improve efficiency, \tool first performs token-based matching,
which consists of three steps: preprocessing, similarity calculation, and pruning.

\begin{itemize}
    \item Preprocessing: Operators, parentheses, and semicolons are removed,
    and identifiers, keywords, and digits are splitted by camel case to
    extract lowercase tokens from the context.
    For example, \texttt{getUnboundParameters()} is tokenized into \texttt{[get, unbound, parameters]}.
    \item Similarity Calculation: Each context including the target statement 
    is treated as a document. \tool then computes TF-IDF similarity scores
    to measure token-level similarity between locations.
    As a lightweight and computationally efficient method,
    TF-IDF is well suited for large-scale location filtering.
    \item Pruning: The top 100 most similar locations (contexts) are retained as 
    initial candidate siblings. 
    We use 100 as it provides a good balance between
    efficiency and effectiveness by preserving relevant candidates 
    while substantially reducing the search space.
    For example, in Math\_100, this step reduces the candidate set by 75\%
    while retaining the relevant candidates,
    reducing the subsequent embedding computation time from hours to less than 5 minutes.
\end{itemize}

\subsubsection{Embedding-based Matching}
\tool next applies embedding-based code matching to identify
semantics-related locations as the candidate siblings 
for subsequent simultaneous and iterative repair.
To this end, it uses the TEXT-EMBEDDING-ADA-002 model~\cite{greene2022embedding}
to generate a semantic embedding vector for each location (context).
It then computes the cosine similarity between
each candidate embedding and that of the suspicious location.
Finally, a predefined similarity threshold is applied to select the final sibling candidates.

We chose TEXT-EMBEDDING-ADA-002 for its strong cross-domain robustness
and its independence from language-specific structural assumptions. 
In contrast, code embedding models such as Code2Vec~\cite{alon2019code2vec}
are primarily designed for method-level tasks (e.g., method name prediction) and 
optimized to capture structural patterns of entire functions
through AST-based representations. Our task however focuses on
semantic similarity across code contexts rather than method-level summarization or structural modeling. 
Moreover, prior work~\cite{kang2019Assessing} suggests that 
source-code token embeddings do not generalize well to downstream tasks.

\begin{algorithm}
\footnotesize
\caption{Simultaneous Repair Algorithm 
}
\label{alg:simrepair}
\KwInput{list of candidate siblings with contexts ${L_{\text{ctx}\_\text{rel}}}$, 
initial buggy program $b$, test suite $T$, 
promising patches $P_{pro}$, plausible patches $P_{pl}$,
Jaccard similarity threshold $\alpha$, repair attempts $t$, max fix ingredients of each line $n$}
\KwOutput{update $P_{pl}$}
\Fn{\textsc{SimRepair}(${L_{\text{ctx}\_\text{rel}}}, b, t, \alpha, n, P_{pro}, P_{pl}, T$)}{
${L^{\prime}_{\text{ctx}\_\text{rel}}} \gets \textsc{JaccardFilter}({L_{\text{ctx}\_\text{rel}}}, \alpha)$\;
$M_{rel}' \gets \textsc{GroupByMethod}({L^{\prime}_{\text{ctx}\_\text{rel}}})$\;
$fb \gets [\;]$;     \tcp{initial feedback info}
$tried \gets 0$;  \tcp{initial attempts without promising patch}
\While{$tried < t$}{
    $prompt \gets \textsc{ConstructPrompt}(M_{rel}', b, T, fb, n)$\;
    $p \gets \textsc{GeneratePatch}(prompt)$\;
    $result \gets \textsc{validate}(p, b, T)$\;
    \eIf{$\text{result} = \text{``pass all''}$}{
        $P_{pl}.\text{add}(p)$\;
        $fb \gets [\;]$\;
        \textsc{Append}$(fb, p)$;
    }{
        $fb \gets [\;]$\;
        $\textsc{Append}(fb, (p, \textsc{GetTestResults}(p, b, T)))$\;
    }
    $tried \gets tried + 1$\;
}
\For{$p_{pro} \;\mathrm{in}\; P_{pro}$}{
    $fb \gets [\;]$\;
    $\textsc{Append}(fb, (p_{pro}, \textsc{GetTestResults}(p_{pro}, b, T)))$\;
    $tried \gets 0$\;
    \While{$tried < t$}{
        $prompt \gets \textsc{ConstructPrompt}(M_{rel}', b, T, fb, n)$\;
        $p \gets \textsc{GeneratePatch}(prompt) + p_{pro}$\;
        $result \gets \textsc{validate}(p, b, T)$\;
        \eIf{$\text{result} = \text{``pass all''}$}{
            $P_{pl}.\text{add}(p)$\;
            $fb \gets [\;]$\;
            \textsc{Append}$(fb, p)$;
        }{
            $fb \gets [\;]$\;
            \textsc{Append}$(fb, (p, \textsc{GetTestResults}(p, b, T)))$;
        }
        $tried \gets tried + 1$\;
    }
}
}
\end{algorithm}

\subsection{Simultaneous Repair}
\begin{figure}[t]
    \centering
    \includegraphics[width=\linewidth]{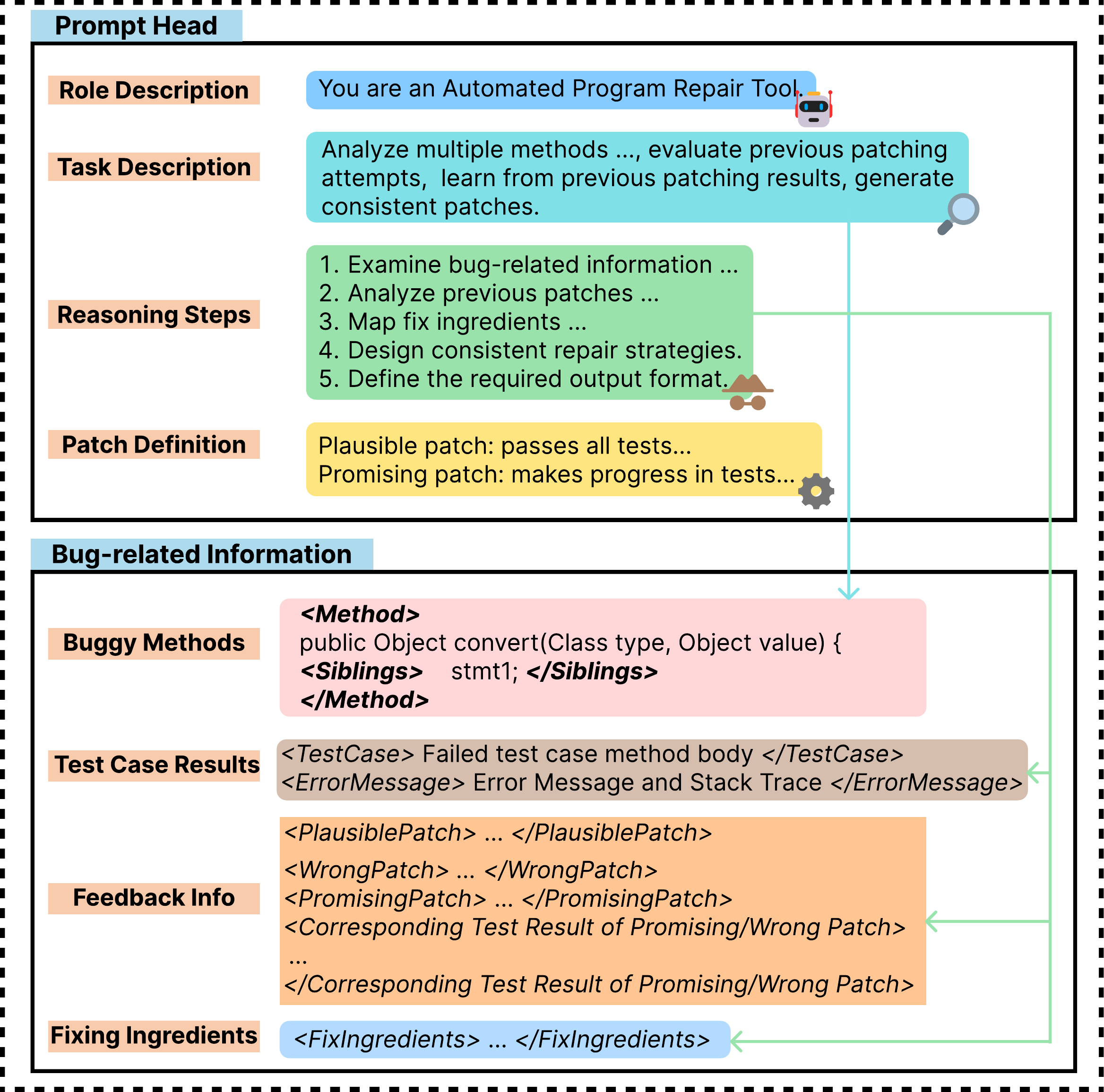}
    \caption{Illustration of the prompt construction. 
    }
    \label{fig:prompt}
\end{figure}

In this step, \tool jointly identifies the failure-relevant locations
from the candidate siblings and generates patches for them, with the help of the LLM.
Algorithm~\ref{alg:simrepair} details the process. 
Given the candidate siblings identified in the previous step, 
\tool first filters out those with low similarity (line~2), 
as such candidates are unlikely to share similar issues with the target. 
Here \tool focuses on lexical similarity and uses Jaccard similarity scores.
This step helps make the remaining candidates fit within the repair prompt. 
It then groups the remaining candidates by their enclosing methods (line~3), 
marking the sibling locations within each method.
Following prior work~\cite{chen2025studying}, patch generation is performed at the method level.

\tool first attempts independent repair without using prior 
promising patches carried over from previous suspicious location repair (lines 6-19).
It then performs repair by considering each promising patch in turn (lines 20-38).
In both cases, it constructs a prompt with five components to
guide the LLM in generating similar patches (lines 7-8 \& 25-26).
Plausible patches that cause all tests to pass are retained
not only because they resolve the observed failure (lines 11 \& 29),
but also because they serve as valuable guidance for the LLM to generate more diverse candidate patches,
thereby increasing the likelihood of producing a correct fix.
Meanwhile, failing patches, along with their associated test feedback,
are also preserved (lines 15-16 \& 33-34) to support subsequent patch refinement.

\begin{algorithm}
\footnotesize
\caption{Fixing Ingredients Extraction}
\label{alg:fix-ingredient-extract}
\KwInput{enclosing methods of the filtered
siblings $M$, buggy program $b$, max fix ingredients of each statement $n$}
\KwOutput{repair prompt $prompt$}
\Fn{\textsc{FixingIngredientsExtraction}($M, b, n$)}{
$I_{fix} \gets \emptyset$; \tcp{initial fix ingredients}
\For{$m \;\mathrm{in}\; M$}{
    $u \gets \emptyset$; \tcp{initial called methods and fields}
    \For{$\ell \;\mathrm{in}\; m.\textit{sibling\_lines}$}{
        $u \gets u \cup \textsc{CallsandFields}(\ell, m.\textit{file\_path}, b)$\;
        $I_{fix} \gets I_{fix} \cup u$\;
        $C \gets \textsc{ExtractSourceFile}(u, b)$\;
        $D \gets \textsc{ExtractCalls}(C)$\;
        $D' \gets \textsc{TokenMatch}(u, D, n)$\;
        $I_{fix} \gets I_{fix} \cup D'$\;
    }
}
\Return{$I_{fix}$}\;
}
\end{algorithm}

The prompt used by \tool is shown in Fig.~\ref{fig:prompt}.
It consists of four descriptive components in the prompt head,
followed by a structured bug-related information section.
The latter includes the buggy methods, test case results,
feedback information, and fixing ingredients. 

\subsubsection{Role Description} The prompt begins with a role specification: 
``You are an Automated Program Repair Tool.'' 
This instruction aligns the LLM with the repair objective.

\subsubsection{Task Description} This component provides a high-level description of the repair task,
instructing the LLM to (a) analyze multiple methods together with 
their labeled sibling statements, (b) evaluate previous patching attempts,
(c) learn from previous patching results,
and (d) generate consistent patches.

\subsubsection{Reasoning Steps} 
This component guides the LLM through the repair process using the following steps:
\begin{itemize}
    \item Examine bug-related information and identify failure-relevant siblings;
    \item Analyze previous patches and infer repair rationales;
    \item Map fixing ingredients to the inferred repair rationales;
    \item Design consistent repair strategies; 
    \item Define the required output format.
\end{itemize}
In steps 1–3, \tool assembles bug-related information,
the previous patches with feedback, and fixing ingredients.
In step 4, \tool requests that LLM generate consistent patches for the siblings.
Finally, in step 5, \tool defines the output format.

In step 1, to collect bug-related information,
\tool includes the failing test cases, runtime error messages, and stack traces,
which help the LLM understand failure symptoms and potential root causes.
In step 2, to allow the LLM to reuse previous patches,
\tool includes the patches and their corresponding test outcomes 
as feedback into the prompt.
The previous patches can be plausible patches, incorrect patches, or promising patches.
The plausible patches allow the LLM to explore alternative consistent repairs 
so as to increase patch generation diversity; 
the incorrect patches instruct the LLM to refine its previous failing repair attempts;
and the promising patches (from previous suspicious location repair)
helps construct general multi-hunk patches.

To collect fixing ingredients in step 3,
\tool uses Algorithm~\ref{alg:fix-ingredient-extract}.
For each enclosing method of the filtered siblings,
\tool first extracts all fields and methods referenced at the sibling locations (lines~6-7), 
then identifies their enclosing classes (line~8), and collects all fields and
method declarations from those classes (line~9).
These elements are ranked by token-level similarity (TF-IDF),
and the top $n$ items are selected as fixing ingredients (lines~10-11).

\subsubsection{Patch Definitions} This part shows what plausible and promising patches are.
A plausible patch is described as a patch that makes all tests pass.
A promising patch is one that partially addresses the root cause and
leads to repair progress in terms of test outcomes and execution.

\subsection{Iterative Repair}
If simultaneous repair fails due to
sibling filtering errors or LLM's inability to correctly identify and repair siblings,
\tool performs iterative repair, as detailed in Algorithm~\ref{alg:iterrepair}.

In iterative repair, \tool first groups candidate siblings by their enclosing methods (line~2)
and then examines these methods one at a time for patch generation (line~3).
For each candidate method, following Algorithm~\ref{alg:simrepair},
it performs both independent repair (lines~7-24), which ignores previously identified promising patches,
and dependent repair (lines~25-49),
which incorporates such patches accumulated from prior candidate methods or suspicious locations.
Independent repair helps avoid error propagation from earlier iterations,
including mistakes in fault localization or sibling identification,
whereas dependent repair enables the construction of combined patches for multi-hunk fixes.
In both cases, \tool prompts the LLM to generate patches only for the current method and
validates them against the test suite to obtain feedback (lines~8-11 \& 30-33).
If a generated patch is plausible, it is stored in $P_{pl}$. 
Otherwise, \tool evaluates whether the patch is promising; 
if so, it is first stored in $newPro$, and then merged into $P_{pro}$
for use across subsequent suspicious locations. 
Candidate methods that yield no promising patches are skipped.

To determine whether a patch is promising,
\tool evaluates repair progress, as reflected by a reduction of failing tests or extended execution progress.
Specifically, a patch is considered promising if it causes any previously failing test to pass.
If no previously failing test is observed to pass, \tool further checks
whether the patch enables execution to proceed beyond a previously observed failure point---such as an assertion failure or an exception-throwing location---for any failing test.
To this end, \tool compares the stack traces obtained before and after applying the patch. 
More specifically, it aligns the stack frames
from the test case downward until the point of divergence,
and then checks whether the patched execution terminates 
within the same method but at a later location, as determined by the line number.
When the diverging frames correspond to different methods 
while all preceding frames remain identical, \tool still treats the patch as promising, 
as such a behavioral change may still indicate repair progress through altered control flow.
Although stack-trace-based comparison may introduce false positives,
we found it to be empirically effective.
A more precise alternative would be to
instrument the program and compare full execution traces; however, such an approach 
incurs substantially higher overhead due to the length and granularity of execution traces. 
We leave an efficient execution-trace-based promising patch detection to future work.

\begin{algorithm}
\footnotesize
\caption{
Iterative Repair Algorithm 
}\label{alg:iterrepair}
\KwInput{list of candidate siblings with contexts ${L_{\text{ctx}\_\text{rel}}}$, initial buggy program $b$, test suite $T$, promising patches $P_{pro}$, plausible patches $P_{pl}$, repair attempts $t$, max fix ingredients of each line $n$}
\KwOutput{Updated $P_{pro}$ and $P_{pl}$}
\Fn{\textsc{IterRepair}($L_{\text{ctx}\_\text{rel}}, b, t, n, P_{pro}, P_{pl}, T$)}{
$M_{rel} \gets \textsc{GroupByMethod}({L_{\text{ctx}\_\text{rel}}})$\;
\For{$m \;\mathrm{in}\; M_{rel}$}{
$fb \gets [\;]$;      \tcp{initial feedback info}
$newPro \gets \emptyset$;  \tcp{initial new promising patches}
$tried \gets 0$;           \tcp{initial repair attempts}
\While{$tried < t$}{
    $M \gets \{m\}$;  \tcp{wrap $m$ as a singleton method set}
    $prompt \gets \textsc{ConstructPrompt}(M, b, T, fb, n)$\;
    $p \gets \textsc{GeneratePatch}(prompt)$\;
    $result \gets \textsc{validate}(p, b, T)$\;
    \eIf{$\text{result} = \text{``pass all''}$}{
        $P_{pl}.\text{add}(p)$\;
        $fb \gets [\;]$\;
        \textsc{Append}$(fb, p)$;
    }
    {
        \If{$\text{result} = \text{``better''}$}{
            $newPro.\text{add}(p)$\;
        }
        $fb \gets [\;]$\;
        $\textsc{Append}(fb, (p, \textsc{GetTestResults}(p, b, T)))$\;
    }
    $tried \gets tried + 1$\;
}
\For{$p_{pro} \;\mathrm{in}\; P_{pro}$}{
    $fb \gets [\;]$\;
    $\textsc{Append}(fb, (p_{pro}, \textsc{GetTestResults}(p_{pro}, b, T)))$\;
    $tried \gets 0$\;
    \While{$tried < t$}{
        $M \gets \{m\}$;  \tcp{wrap $m$ as a singleton method set}
        $prompt \gets \textsc{ConstructPrompt}(M, b, T, fb, n)$\;
        $p \gets \textsc{GeneratePatch}(prompt) + p_{pro}$\;
        $result \gets \textsc{validate}(p, b, T)$\;
        \eIf{$\text{result} = \text{``pass all''}$}{
            $P_{pl}.\text{add}(p)$\;
            $fb \gets [\;]$\;
            \textsc{Append}$(fb, p)$;
        }
        {
            \eIf{$\text{result} = \text{``better''}$}{
                $newPro.\text{add}(p)$\;
            }
            {
                $newPro.\text{add}(p_{pro})$\;
            }
            $fb \gets [\;]$\;
            $\textsc{Append}(fb, (p, \textsc{GetTestResults}(p, b, T)))$\;
        }
        $tried \gets tried + 1$\;
    }
}
$P_{pro} \gets newPro$\;
}
}
\end{algorithm}

\section{Experimental Settings}
\label{sec:ex-settings}

To assess the effectiveness of \tool, we compare it with
state-of-the-art multi-hunk APR techniques on
the Defects4J v2.0.1 benchmark~\cite{defects4jrepo} under
two fault localization settings: spectrum-based fault localization (SBFL) and
single perfect fault localization (SPFL).

Under SBFL, repair techniques use spectrum-based fault localization with
the Ochiai metric to obtain a ranked list of suspicious locations for repair.
The setting reflects realistic end-to-end APR usage.
Under SPFL, one real buggy location $l_k$ is assumed to be known.
Among all real buggy locations derived from the developer patch,
$l_k$ denotes the location with the highest suspiciousness score.
For the repair experiments, the original suspicious list is modified 
such that $l_k$ becomes the top-ranked suspicious location, 
while all other locations retain their relative order with slightly lower ranks.
This modified list is then used for repair.
We conducted SPFL experiments to better understand the repair abilities of techniques
such as \tool and \hercules, which initiate multi-hunk repair from a single suspicious location.

To further evaluate \tool's ability to repair sibling bugs,
we also conduct experiments on the sibling bugs from GHRB~\cite{lee2024github}.
We include GHRB in addition to Defects4J
because it contains more recent bugs and presents a lower risk of data leakage for LLM-based evaluation.
For Defects4J, \tool is configured to use DeepSeek-V3.2.
For GHRB, it is evaluated with both DeepSeek-V3.2 and GPT-3.5-Turbo,
whose training cutoff predates the GHRB bugs.

We designed five research questions (RQs). 
Next, we present each RQ together with its corresponding experimental setup in detail.

\subsection{Research Questions}

\textbf{RQ-1: How does \tool perform under SBFL?}

To answer this question, we evaluate \tool under the realistic SBFL setting 
and compare it against APR techniques that are most relevant to our scenario,
namely sibling-based or general multi-hunk repair approaches operating on 
SBFL-generated suspicious lists rather than assuming perfect fault localization.
Following this criterion, we include representative approaches
for multi-hunk repair across different paradigms,
including \hercules (sibling-oriented template-based repair)~\cite{saha2019harnessing},
ITER (iterative learning-based repair)~\cite{ye2024iter}, and ARJACLM
(search-based repair with LLM-assisted ingredients)~\cite{lijzenga2025leveraging}.
These baselines are closely aligned with our focus while covering diverse repair strategies. 
For these techniques, we used their publicly available implementations 
at~\cite{herculesreimplemented, iter, arjaclm}
with only minimal adaptations required for execution compatibility with our experimental environment.

We exclude several APR techniques that are not well aligned with our setting.
First, search-based multi-hunk approaches such as
GenProg~\cite{le2011genprog}, ARJA~\cite{yuan2018arja}, and ARJA-e~\cite{yuan2019hybrid}
are not included because prior evaluations have shown them to be inferior to ARJACLM,
which serves as a stronger baseline with the same paradigm.
Second, approaches such as MultiMend~\cite{gharibi2025multimend},
RepairAgent~\cite{bouzenia2024repairagent}\footnote{Although the authors state that 
the approach can be combined with SBFL, they do not describe how this integration 
supports multi-hunk repair, nor do they provide an implementation that incorporates SBFL.}, RepairLLaMA~\cite{silva2025repairllama}, PReMM~\cite{xie2025premm},
DynaFix-v1~\cite{huang2025dynafix}\footnote{While DynaFix-v1~\cite{huang2025dynafix} 
assumes PFL, DynaFix-v2~\cite{huang2026dynafix} additionally supports the SBFL configuration 
and is thus comparable to \tool.
However, DynaFix-v2 was released only recently (on April 19th) and
was unavailable at the time of our experiments, so we did not include it in our evaluation.
Nevertheless, according to its reported results on Defects4J-v1.2
(no results on v2.0), DynaFix-v2 repairs 26 fewer bugs than \tool,
indicating weaker effectiveness.},
and SRepair~\cite{xiang2024far} are excluded because they assume perfect fault localization (PFL),
where all buggy locations are known in advance.
This is a strong assumption for multi-hunk repair, making a direct comparison with \tool unfair.
Third, ThinkRepair~\cite{yin2024thinkrepair} is not included,
as the approach assumes a known location (i.e., a buggy method).
We nevertheless compare \tool and ThinkRepair under SPFL (see RQ2).
Finally, we do not consider non-test-based repair techniques
such as AutoCodeRover~\cite{AutoCodeRover}, SpecRover~\cite{SpecRover},
and Agentless~\cite{Agentless}
or single-hunk repair techniques including FitRepair~\cite{xia2023revisiting},
Gamma~\cite{zhang2023gamma}, and AlphaRepair~\cite{xia2022less}, 
since our primary focus is multi-hunk repair under test-based scenarios.

We ran \tool, \hercules, and ARJACLM on all Defects4J bugs
to obtain their repair results. For ITER, we used the results
from its original evaluation for 501 bugs from 10 Defects4J projects.
For the remaining 334 bugs, for which no results were reported,
we ran the tool to obtain the missing results. Following~\cite{xin2024detecting}, 
we use a 5-hour time budget for repair. 

\textbf{RQ-2: How does \tool perform under SPFL?}

RQ-2 complements RQ-1 by focusing on multi-hunk repair starting from a single fault location, 
which aligns with the repair paradigm of sibling-based techniques such as \tool.
We evaluate all four tools---\tool, \hercules, ARJACLM, and ITER---under SPFL,
using the modified suspicious list, where a real buggy location
is ranked first, as described at the beginning of Section~\ref{sec:ex-settings}.
We run these tools to obtain the results under SPFL, using a 5-hour time budget.
ThinkRepair, in contrast, is designed to handle only multi-hunk bugs within
a single method and assumes that the buggy method is given, 
rather than operating over a suspicious list. This difference makes the comparison 
not strictly equivalent. Nevertheless, to provide a point of reference, 
we reuse its previously reported results under the assumption that the buggy method is known in advance.

\textbf{RQ-3: What is the repair time cost of \tool?}

In this RQ, we investigate \tool's efficiency by measuring the time required to repair bugs.
Specifically, we consider the elapsed time from the start of the candidate sibling detection phase
to the end of the repair process. The process terminates either when
plausible patches are generated or when the 5-hour time budget is exhausted.
This measurement reflects the total repair time of \tool, excluding the fault localization phase.

\textbf{RQ-4: What is the effectiveness of each component in \tool?}

In RQ-4, we evaluate the effectiveness of \tool's key components:
candidate sibling detection, simultaneous repair, and iterative repair. 
Specifically, we first assess the recall of candidate sibling detection
to determine whether the candidate set includes the failure-relevant siblings.
We focus on recall rather than precision at this stage because the subsequent repair stages
are responsible for identifying the true failure-relevant siblings.
We then analyze the contributions of the simultaneous repair and 
iterative repair stages in terms of their impact on overall repair performance.

Rather than performing a standard ablation study, we conduct 
an effectiveness analysis, as these components operate in a sequential and interdependent pipeline.
Candidate sibling detection provides the input to the subsequent repair stages, 
while iterative repair builds upon the outcomes of simultaneous repair.
Under this design, isolating individual components through standard ablation 
is less meaningful, since removing an earlier stage would invalidate 
the inputs required by later stages.

\textbf{RQ-5: How does \tool perform on additional recent bugs?}

To further assess the generalizability of \tool, we additionally evaluate it 
on the sibling bugs from the GHRB benchmark,
as introduced at the beginning of Section~\ref{sec:ex-settings}.
To identify sibling bugs, we examine the developer patches of all GHRB bugs
provided in the benchmark and determine whether the multi-hunk changes
across different locations are similar. Specifically, we first extract
hunk-level modifications using GumTree~\cite{falleri24fine}, and then compute
pairwise edit-text similarity using Python's \texttt{difflib.SequenceMatcher}.
We set the similarity threshold as 0.6 and classify a bug as sibling-based
only if each hunk modification has at least one other hunk modification with similarity no lower than 0.6.
This threshold is chosen based on empirical observation.
Using this procedure, we identify 17 siblings bugs for evaluation.
We also apply the same method to Defects4J and identify 77 sibling bugs
for analysis purposes.
For the GHRB bugs, \tool is run with two LLM models, DeepSeek-V3.2 and GPT-3.5-Turbo.
We include GPT-3.5-Turbo because its training cutoff predates these bugs, 
providing a stricter setting for controlling potential data leakage.

\subsection{Patch Assessment}
\label{sec:patch_assessment}

All plausible patches are manually inspected against the developer patch
(ground truth) to determine repair correctness.
Specifically, a patch is considered correct if it is either syntactically identical
or semantically equivalent to the corresponding developer patch.
Given the total number of plausible patches (1,751), we divide them into
three subsets and assign each subset to a different author for independent assessment.
For the majority of patches (1,529), the assigned author is able to make a confident judgment,
as the distinction between correct and incorrect patches is typically straightforward.
In these cases, incorrect patches either introduce
substantially different modifications from the developer patch,
or resemble it while omitting necessary changes.
By contrast, correct patches are highly similar and semantically equivalent to the developer patch. 
For the remaining 222 patches where the assigned author is uncertain, 
all three authors independently review these patches and determine their correctness 
through discussion and consensus.

\subsection{Experimental Environment}
All experiments were conducted on machines with 16 Intel i9-11900K 3.5 GHz CPUs, 
NVIDIA RTX A4000/5000 GPUs of 16/24 GB, and 128 GB RAM, running Ubuntu-20.04.

\section{Experimental Results}

\begin{table}[t]
\caption{Plausible and correct repairs on Defects4J under SBFL.
}
\begin{threeparttable}
\footnotesize
\setlength{\tabcolsep}{4pt}  
\resizebox{\columnwidth}{!}{
\begin{tabular}{lrrrr}
\toprule
Dataset                   & \tool   &     \hercules  &  ARJACLM   &    ITER    \ignore{ &  ThinkRepair }\\
\midrule
Chart                     & 10/20   &      5/10     &     4/7    &    10/14   \ignore{ &   9/-  } \\
Closure                   & 18/49   &      1/1      &     2/11   &    17/22   \ignore{ &   19/- } \\
Lang                      & 21/57   &      6/12     &     6/11   &    9/11    \ignore{ &   15/- } \\
Math                      & 31/70   &      6/14     &     7/27   &    20/36   \ignore{ &   27/- } \\
Mockito                   & 11/22   &      0/1      &     1/3    &    0/0     \ignore{ &   7/-  } \\
Time                      & 3/7     &      0/2      &     1/1    &    2/4     \ignore{ &   3/-  } \\
\midrule
Defects4J 1.2             & 94/225   &     18/40    &     21/60  &    58/87   \ignore{ &   80/- } \\
\midrule
Cli                       & 1/4      &     1/8      &     3/5    &    6/13    \ignore{ &   -/-  } \\
Closure                   & 3/7      &     0/0      &     1/4    &    0/0     \ignore{ &   -/-  } \\
Codec                     & 9/14     &     1/6      &     0/5    &    3/6     \ignore{ &   -/-  } \\
Collections               & 0/3      &     0/0      &     0/1    &    0/0     \ignore{ &   -/-  } \\
Compress                  & 10/27    &     1/10     &     3/8    &    4/6     \ignore{ &   -/-  } \\
Csv                       & 4/13     &     3/4      &     2/2    &    2/5     \ignore{ &   -/-  } \\
Gson                      & 6/11     &     1/2      &     1/2    &    0/0     \ignore{ &   -/-  } \\
JacksonCore               & 6/15     &     0/6      &     0/8    &    3/6     \ignore{ &   -/-  } \\
JacksonDatabind           & 12/61    &     0/5      &     1/19   &    1/7     \ignore{ &   -/-  } \\
JacksonXml                & 4/4      &     0/0      &     0/2    &    0/0     \ignore{ &   -/-  } \\
Jsoup                     & 23/60    &     3/6      &     4/12   &    1/6     \ignore{ &   -/-  } \\
JxPath                    & 4/10     &     0/4      &     1/4    &    0/2     \ignore{ &   -/-  } \\
\midrule
Defects4J 2.0             & 82/229   &     10/51    &     16/72  &   20/51    \ignore{ &  90/-  } \\
\midrule
Total                     & 176/454  &     28/91    &     37/132 &   78/138   \ignore{ &  170/- } \\
\bottomrule
\end{tabular}
}
\vspace{3pt}
\begin{tablenotes}[flushleft]
     \footnotesize
     \item[*] \parbox{0.97\columnwidth}{x/y: x and y are the numbers of bugs for which correct patches and plausible patches were generated respectively.}
 \end{tablenotes}
\end{threeparttable}

\label{tab:fl}
\end{table}

\subsection{RQ1: How does \tool perform under SBFL?}
Table~\ref{tab:fl} presents the repair results of Defects4J by projects,
and Table~\ref{tab:hunk} shows correct repairs of
single-hunk (SH), multi-hunk (MH), and sibling (Sibling) bugs.
Under SBFL in Table~\ref{tab:hunk}, \tool repairs 23 sibling bugs
(15 from v1.2 and 8 from v2.0), substantially outperforming its sibling repair counterpart \hercules, 
which repairs only 4 sibling bugs.
Even as a sibling-oriented technique, \tool repairs 76 multi-hunk bugs 
(44 from v1.2 and 32 from v2.0), with the best baseline ITER repairing only 24 multi-hunk bugs.
These results demonstrate \tool's superior sibling and multi-hunk repair abilities.
As shown in Table~\ref{tab:fl}, \tool generates correct patches for 176 out of 835 bugs.
It outperforms ITER, ARJACLM, and \hercules, which repair 
78, 37, and 28 bugs respectively.

\begin{tcolorbox}[
    colback=white,
    colframe=black,
    boxrule=0.5pt,
    arc=2pt,
    left=4pt,right=4pt,top=4pt,bottom=4pt
]
\textbf{Answer to RQ-1:} 
\textit{
Under SBFL, \tool outperforms \hercules, ARJACLM, and ITER on Defects4J,
achieving the highest number of correct repairs. 
It is particularly effective on sibling and multi-hunk bugs, 
repairing about 6x as many sibling bugs as \hercules and
about 3x as many multi-hunk bugs as ITER.
}
\end{tcolorbox}

\subsection{RQ2: How does \tool perform under SPFL?}
\label{sec:rq2}

Under SPFL, where a real buggy location is used as the repair starting point,
\tool repairs many more bugs, as shown in Table~\ref{tab:hunk}.
Specifically, it correctly repairs 29 sibling bugs (19 from v1.2 and 10 from v2.0),
whereas \hercules repairs only 4. For multi-hunk bugs, it repairs 124 (71 from v1.2 and 53 from v2.0), 
while the strongest baseline ThinkRepair repairs only 37.
These results demonstrate \tool's strong capability to perform repair
starting from a single location, particularly for sibling bugs and more broadly multi-hunk bugs.
This makes it well suited to current fault-localization techniques that
provide a ranked list of suspicious locations.
These results also suggest strong potential of \tool for further 
improving multi-hunk repair performance as the accuracy of the underlying fault localization improves.

Of the 77 sibling bugs, \tool correctly repairs 29 and fails on the remaining 48 bugs.
We analyze the failures and find that 18 are caused by
incomplete test coverage, where the test suite does not exercise
all failure-relevant siblings, preventing \tool from including them into its repair scope.
Among the remaining 30 bugs, 12 failures are due to candidate sibling detection,
where the token- and embedding-based matching fails to recognize the 
failure-relevant siblings as sufficiently related.
One possible mitigation is to lower the similarity threshold to retain more relevant locations.
However, this would also introduce more irrelevant candidates,
enlarging the search space and potentially reducing the accuracy of subsequent
LLM-based sibling identification and patch generation. 
For the remaining 18 bugs, all failure-relevant siblings are included in the candidate set,
and the bottleneck shifts to LLM-based repair stages. 
Specifically, the LLM fails to select some failure-relevant siblings for 4 bugs 
and incorrectly selects irrelevant locations for 3 bugs.
Our analysis suggests that these cases often involve complex execution behavior, 
where the true sibling locations are deeply embedded in the execution trace. 
Without sufficient dynamic execution context, the LLM may struggle to 
make accurate sibling selections. Finally, for the remaining 11 bugs, 
the LLM still fails to generate correct repairs even when the failure-relevant siblings 
are correctly selected. These failures are primarily due to hallucinations 
during simultaneous patch construction and the inability of iterative repair to 
identify promising patches when no clear repair progress is observed.

\begin{tcolorbox}[
    colback=white,
    colframe=black,
    boxrule=0.5pt,
    arc=2pt,
    left=4pt,right=4pt,top=4pt,bottom=4pt
]
\textbf{Answer to RQ-2:} 
\textit{Under SPFL, \tool demonstrates strong multi-hunk repair capability
when starting from a single suspicious location, 
substantially outperforming all baselines, particularly on sibling and multi-hunk bugs; 
its failures stem from incomplete test coverage, missed sibling detection,
limitations of LLM-based sibling selection and patch generation, and
progress identification for iterative repair.}

\end{tcolorbox}

\begin{table}[t]
\centering
\caption{Correct repair of Defects4J single-hunk, multi-hunk, and sibling bugs under SBFL and SPFL.}
\begin{threeparttable}
\scriptsize
\setlength{\tabcolsep}{2.5pt} 

\resizebox{\columnwidth}{!}{%
\begin{tabular}{l|l|cccc|cccc|c}

\toprule
\multirow[c]{3}{*}[-1.5ex]{\makecell[c]{Setting}} & \multirow[c]{3}{*}[-1.5ex]{\makecell[c]{Tool}}
& \multicolumn{9}{c}{Defects4J} \\
\cmidrule(lr){3-11}
& & \multicolumn{4}{c}{V1.2}
& \multicolumn{4}{c}{V2.0}
& \multirow{2}{*}{All} \\
\cmidrule(lr){3-6}\cmidrule(lr){7-10}
& & SH & MH & Sibling & Total
& SH & MH & Sibling & Total
& Total\\
\midrule
\multirow{4}{*}{\makecell[c]{SBFL}}
& \hercules
& 13 & 5 & 4 & 18 
& 6 & 4 & 0 & 10 & 28\\ 
& ARJACLM
& 15  & 6  & 3 & 21
& 13  & 3  & 0 & 16 & 37 \\
& ITER
& 39 & 19 & 10 & 58
& 15 & 5 & 3 & 20 & 78\\

& \textbf{\tool}
& \textbf{50} & \textbf{44} & \textbf{15} & \textbf{94} 
& \textbf{50} & \textbf{32} & \textbf{8} & \textbf{82} & \textbf{176}\\ 

\midrule
\multirow{5}{*}{\makecell[c]{SPFL}}
& \hercules
& 14 & 5 & 4 & 19   
& 9 & 4 & 0 & 13 & 32\\ 
& ARJACLM
& 16  & 7  & 2 & 23
& 14  & 3  & 0 & 17 & 40 \\
& ITER
& 14 & 5 & 3 & 19
& 11 & 2 & 0 & 13 & 32\\

& ThinkRepair
& 64  & 16 & - & 80 
& 69  & 21 & - & 90 & 170 \\ 

& \textbf{\tool}
& \textbf{81} & \textbf{71}   & \textbf{19} & \textbf{152}  
& \textbf{73} & \textbf{53}  & \textbf{10} & \textbf{126} & \textbf{278} \\ 

\bottomrule
\end{tabular}%
}
\begin{tablenotes}
    \footnotesize
     \item[*] SH: Single-hunk bugs; MH: Multi-hunk bugs; 
     Sibling: Sibling bugs (a special type of multi-hunk bugs); Total: SH + MH. 
     Since ThinkRepair assumes a known location (a buggy method),
     we only compare it with other tools under SPFL.
     We run the ITER tool provided by the authors~\cite{iter} for bug repair under SPFL.
     In this setting, ITER does not achieve better repair than under SBFL,
     which includes results from the previous evaluation.
\end{tablenotes}
\end{threeparttable}
\label{tab:hunk}
\end{table}

\subsection{RQ-3: What is the repair time cost of \tool?}

\begin{figure}[t]
    \centering
    \includegraphics[width=\linewidth]{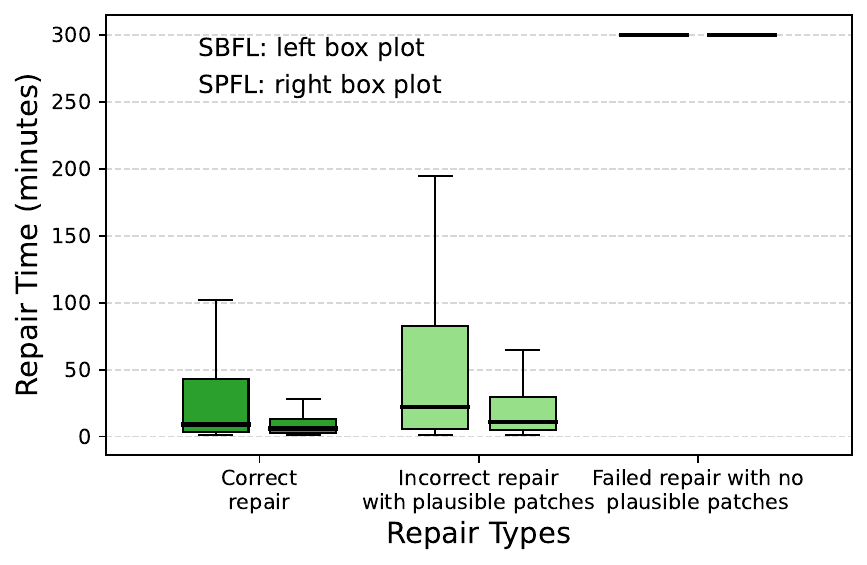}
    \caption{Box plots of \tool's repair time for the Defects4J bugs
    under SBFL (left) and SPFL (right) for each setting. Time for fault localization is excluded.
    From left to right, it shows the statistics for bugs correctly repaired,
    incorrectly repaired with plausible patches, and not repaired with no plausible patches.
    }
    \label{fig:repair_time}
\end{figure}

Fig.~\ref{fig:repair_time} presents the time cost of \tool's repair process,
excluding fault localization. The box plots summarize the distribution of 
repair time across all the Defects4J bugs under SBFL and SPFL.
Since fault localization is excluded, the difference between the two settings 
arises from the suspicious location lists used.
Recall that under SBFL, the list is ranked using the Ochiai metric,
whereas under SPFL, it is modified by placing a real buggy location at the top.
Fig.~\ref{fig:repair_time} reports results for three categories of bugs:
those correctly repaired (left), those incorrectly repaired with plausible patches (middle), 
and those for which no plausible patches are found (right).

The results indicate that \tool is efficient at generating plausible patches.
For correctly repaired bugs under SBFL, the median time for plausible patch generation
is 9 minutes, while for incorrectly repaired bugs with plausible patches,
the median time is 23.5 minutes. Under SPFL, where a real buggy location is examined first,
\tool is faster, requiring a median of 6 minutes to correctly repair a bug.
These findings suggest that \tool can provide timely repair results
if it can find a plausible patch. For bugs that \tool fails with 
no plausible patches, it is almost always the case that \tool uses up the time budget 
but finds no plausible patch.

\begin{tcolorbox}[
    colback=white,
    colframe=black,
    boxrule=0.5pt,
    arc=2pt,
    left=4pt,right=4pt,top=4pt,bottom=4pt
]
\textbf{Answer to RQ-3:} \textit{\tool generates plausible patches efficiently, 
typically within minutes. When it successfully produces a correct patch, 
it usually does so quickly—especially under SPFL—whereas cases with no plausible patches 
tend to exhaust the full time budget.}
\end{tcolorbox}

\subsection{RQ-4: What is the effectiveness of each component in \tool?}

In this RQ, we evaluate the effectiveness of \tool's core components---candidate sibling detection,
simultaneous repair, and iterative repair (applied when simultaneous repair fails)---on the 77 Defects4J sibling bugs.

For candidate sibling detection, we assess recall at both the statement and method levels.
While \tool performs detection at the statement level,
we also report method-level results because the LLM-based repair 
operates at the method level and can tolerate inaccuracies
in statement-level detection within a buggy method.
Our analysis focuses on the SPFL setting, where the initial suspicious location is a real buggy location.
If the initial location is not buggy, missing siblings cannot be attributed to 
weaknesses in candidate sibling detection. 
We focus on recall, as subsequent repair stages further filter failure-relevant siblings 
from the candidates.

The results show that \tool identifies all sibling statements
for 53.2\% (41/77) of the bugs and covers all sibling methods for
70.1\% (54/77) of the bugs. Missed candidate siblings are primarily due to
the limitations of \tool's token- and embedding-based
matching methods in capturing related but less similar code fragments,
as discussed in Section~\ref{sec:rq2}. 

For the 54 bugs where all failure-relevant siblings are included in the candidate set,
we further evaluate simultaneous and iterative repair.
In simultaneous repair, \tool correctly identifies failure-relevant siblings 
for 41 (75.9\%) of these bugs and produces plausible patches for 38 (70.4\%), 
among which 25 (46.3\%) are correct.
For the 13 bugs, the generated patches are plausible but incorrect.
In these cases, the LLM modifies the right methods and resolves the test failures,
but the patches are not strictly semantically consistent with the developer patches,
although sometimes they are also valid.
A representative example is Closure\_131, where the developer patch
additionally invokes \texttt{Character.isIdentifierIgnorable}
to support identifier validation, whereas the LLM instead uses specific characters 
and digits to approximate the legality check.
In this case, the generated patch resolves the immediate failing behavior, 
but it does not precisely match the developer patch because it does not fully 
capture the intended program specification.

When simultaneous repair fails, iterative repair produces plausible patches
for 8 additional bugs, among which 4 are correct.
For the other 4 plausible-but-incorrect cases,
a main issue is that iterative repair produces partial patches 
that make the failing tests (and all tests) pass, revealing
insufficient tests for bug exposing. Note that in these cases 
all the failure-relevant siblings are exercised by at least one test, 
but failing tests are insufficient to expose all of them as buggy.

For the remaining 8 (54-38-8) bugs where \tool does not produce any plausible patch,
the failures are mostly due to bug complexity: even when the true sibling statements
are localized accurately, simultaneous repair still generate incorrect patches,
and iterative repair either fails to produce promising patches or produces 
promising-but-incorrect partial patches that cannot be extended into a correct full repair.

\begin{tcolorbox}[
    colback=white,
    colframe=black,
    boxrule=0.5pt,
    arc=2pt,
    left=4pt,right=4pt,top=4pt,bottom=4pt
]
\textbf{Answer to RQ-4:} \textit{
\tool's components are effective. Candidate sibling detection
achieves 53.2\% recall at the statement level and 70.1\% at the method level.
When all failure-relevant siblings are included, simultaneous repair can correctly
repair about half of the bugs, and iterative repair provides additional benefit.
}
\end{tcolorbox}

\subsection{RQ-5: How does \tool perform on additional recent bugs?} 
To assess the potential impact of LLM's data leakage on the repair performance,
we conduct an experiment in which \tool is applied to the GHRB sibling bugs
using two LLMs: DeepSeek-V3.2 and GPT-3.5-Turbo.
GPT-3.5-Turbo is trained on data up to September 2021,
while GHRB consists of bugs introduced after this date.
As a result, using GPT-3.5-Turbo reduces the risk of data leakage, 
since the model is not exposed to these bugs during training.

In this experiment, we run \tool under perfect fault localization (PFL),
assuming that all buggy locations are known.
This setting is used because the GZoltar tool~\cite{gzoltar_paper}
employed by \tool for fault localization 
does not support the JUnit 5 test cases in GHRB. 
Although APR under PFL is less realistic,
it is appropriate for this RQ, as our goal is to isolate and examine 
the influence of LLM data leakage on repair performance.
Under PFL, buggy locations are ranked according to their order of appearance 
in the developer patches provided in the dataset.

Table~\ref{tab:ghrb} reports the repair results.
Using DeepSeek-V3.2, \tool correctly repairs 7 bugs,
compared to 6 bugs when using GPT-3.5-Turbo.
This difference suggests that data leakage may have some impact 
on repair effectiveness. However, the overall performance remains 
comparable across the two models, indicating that the impact is limited.

\begin{table}[]
\centering
\caption{Plausible and correct repairs on the 17 GHRB sibling bugs
with two LLM models: DeepSeek-V3.2 and GPT-3.5-Turbo under
perfect fault localization (PFL).}

\begin{threeparttable}
\footnotesize
\begin{tabular}{lrrrr}
\toprule
Projects                   & \thead{\tool\\(DeepSeek-V3.2)}  & \thead{\tool\\(GPT-3.5-Turbo)} \\
\midrule
checkstyle                & 5/5    &    5/5  \\
gson                      & 1/1    &    0/0  \\
jackson-core              & 0/0    &    0/1  \\
jackson-databind          & 0/0    &    0/1  \\
jackson-dataformat-xml    & 0/0    &    0/0  \\
nacos                     & 0/0    &    0/0  \\
openapi-generator         & 1/1    &    1/1  \\
rocketmq                  & 0/2    &    0/0  \\
seata                     & 0/0    &    0/0  \\
\midrule
Total                     & 7/9    &    6/8  \\
\bottomrule
\end{tabular}
\begin{tablenotes}
     \footnotesize
     \item[*] x/y: x and y are the numbers of bugs for which correct patches and plausible patches were generated respectively.
 \end{tablenotes}
\end{threeparttable}
\label{tab:ghrb}
\end{table}

\begin{tcolorbox}[
    colback=white,
    colframe=black,
    boxrule=0.5pt,
    arc=2pt,
    left=4pt,right=4pt,top=4pt,bottom=4pt
]
\textbf{Answer to RQ-5:} 
\textit{
While \tool achieves slightly better results with DeepSeek-V3.2, 
which may have been exposed to the GHRB bugs during training, 
its performance with GPT-3.5-Turbo---whose training predates the dataset---remains comparable. 
This suggests that potential data leakage has a limited impact on \tool's effectiveness.
}
\end{tcolorbox}

\section {Threats To Validity}
\label{sec:threats}

\textbf{External threats.}
A potential external threat arises from the data leakage in Defects4J, as many bugs in this benchmark were fixed before the training cutoffs of LLMs and may appear in training data.
To mitigate this threat, we further evaluate \tool using GPT-3.5-Turbo model on GitHub Recent Bugs (GHRB)~\cite{lee2024github}, where bugs are fixed after the training cutoff for GPT-3.5-Turbo: 
September 2021. The results on GHRB show that our approach generalizes beyond Defects4J.

\textbf{Internal threats.}
First, some tools are not available or have errors. The original implementation of HERCULES is not publicly available, so we use the tool re-implemented at~\cite{herculesreimplemented} 
based on the original technical description~\cite{saha2019harnessing}. 
ARJACLM suffers from failed sanity check and bugs in the process of collecting test results.
We fix these errors and use the fixed version~\cite{arjaclmmodified} for experiments. 
Second, we manually assess plausible patches generated by \tool and other baselines. 
To reduce bias, three authors independently review uncertain patches and 
determine their correctness through discussion and consensus.  

\section{Related Work}
In this section, we focus on multi-hunk repair approaches, 
while also reviewing single-hunk repair approaches and related research 
on code clone detection.

\subsection{Multi-hunk Repair}

Multi-hunk repair remains one of the most challenging problems in APR. Although recent advances in deep learning and LLMs have substantially improved the quality of generated patches, many existing learning-based approaches still struggle to generate correct repairs across multiple code locations in real-world settings where the buggy locations are not known a priori.

\hercules~\cite{saha2019harnessing} is the closest work to \tool in 
sibling-based repair but differs in both sibling identification and repair strategy. \hercules identifies siblings from statements covered by failing tests through AST matching and then expands siblings using historical commits. Thus, \hercules struggles with sibling bugs uncovered by failing tests, syntactically different bugs, and early-stage bugs. Instead, \tool uses token-based and embedding-based matching over statements executed by both passing and failing tests to detect candidate repair siblings, and then filters candidates by LLM reasoning.
Considering repair strategy, \hercules applies the same repair template to all siblings, while \tool generates patches with LLM and feedback from failed attempts using two complementary strategies, 
enabling more flexible and effective sibling-based repair.

ITER~\cite{ye2024iter} is a representative APR approach based on iterative repair strategy, which alternates between testing, fault localization, and repair to gradually refine promising patches toward correct patches.
Unlike \tool, ITER does not target sibling bugs. Although related in the iterative part, 
\tool primarily differs from ITER in both identification criteria of promising patches and 
how such patches are reused for effective repair. ITER considers a patch promising if it 
reduces the number of failing tests. In contrast, \tool considers a patch promising if it enables any previous failing test to pass or moves the failure point deeper in the stack trace. 
Moreover, \tool reuses promising patches in both simultaneous and iterative repair processes,
while ITER only adopts iterative repair strategy. These designs allow \tool to better 
explore promising patches and accumulate effective multi-hunk repairs across iterations.

Learning-based APR employs neural network models or LLMs to learn knowledge of program repair from large amounts of data in open source software~\cite{zhang2024systematic} and has shown strong potential in handling multi-hunk bugs~\cite{huang2024evolving}.
ARJACLM~\cite{lijzenga2025leveraging} enhances ARJA~\cite{yuan2018arja} by using LLMs to generate high-quality patch ingredients. MultiMend~\cite{gharibi2025multimend} heuristically combines patches across locations to enable multi-method repair. Other learning-based approaches adopt autonomous agents to diagnose and repair multi-method bugs, such as RepairAgent~\cite{bouzenia2024repairagent}.
In contrast, \tool does not treat LLM outputs merely as auxiliary ingredients, nor does it simply combine partial patches or fully delegate repair decisions to an autonomous agent. Instead, it uses LLMs within a structured framework to identify sibling relationships and generate similar patches for sibling locations, prioritizing consistent fixes across methods.

There are also APR approaches that rely on the assumption of perfect fault localization where the buggy locations are known a priori. These approaches are mainly found in learning-based methods, such as RepairLLaMA~\cite{silva2025repairllama}, SRepair~\cite{xiang2024far}, and PReMM~\cite{xie2025premm}.

\subsection{Single-hunk Repair}

In contrast to multi-hunk repair, a large body of APR research has focused on repairing a single location. Constraint-based approaches, such as SemFix~\cite{nguyen2013semfix}, typically synthesize repairs for an individual buggy location by solving repair constraints. 
Search-based methods narrow the search space by retrieving similar code from the target project and patches mined from software repositories, but they usually support only single-hunk repair, such as ssFix~\cite{xin2017leveraging}, SimFix~\cite{jiang2018shaping} and ConFix~\cite{kim2019automatic}. Template-based approaches generally apply predefined or mined fix patterns to one buggy location at a time, such as ACS~\cite{xiong2017precise} and FixMiner~\cite{koyuncu2020fixminer}. Learning-based approaches have further advanced single-hunk repair by leveraging LLMs or neural network models that are pre-trained to capture repair patterns from large-scale data, including FitRepair~\cite{xia2023revisiting}, Gamma~\cite{zhang2023gamma}, and AlphaRepair~\cite{xia2022less}. 

\subsection{Code Clone Detection}

Our work is related to code clone detection but targets a fundamentally different objective. Code clone detection aims to identify structurally (lexically or syntactically) similar or functionally equivalent code fragments, whereas \tool focuses on identifying code fragments that are likely to require similar modifications, or siblings. Although our approach draws inspiration from token-based and embedding-based techniques commonly used in Type-3 (syntactic similarity) and Type-4 (semantic similarity)~\cite{dou2023towards} clone detection, these techniques are used only to preliminarily identify candidate siblings from suspicious locations. After obtaining candidate siblings, \tool uses an LLM to further reason about true siblings in simultaneous and iterative repair stages. 
We therefore focus on repair-oriented semantic relevance rather than 
structural similarity or functional equivalence.

\section{Conclusion and Future Work}
\tool is an LLM-based multi-hunk APR technique that
targets sibling repair, addressing the common case where similar bugs occur
across multiple locations and require similar fixes.
Unlike prior work, it avoids strong assumptions on test coverage and commit history
by combining token-based filtering with embedding-based semantic matching to
identify candidate siblings across test-exercised code.
It then employs two complementary strategies---simultaneous repair for joint
failure-relevant sibling identification and fixing, and iterative repair
for progressively validating and accumulating promising patches---thus
improving robustness to sibling identification errors and effectiveness of patch generation.
Beyond sibling repair, \tool also supports general multi-hunk repair by 
composing promising patches across locations. 
Our evaluation on Defects4J and GHRB shows that \tool substantially outperforms
state-of-the-art APR techniques in repairing both sibling bugs and general multi-hunk bugs,
demonstrating its advancement in APR.

For future work, we plan to explore richer dynamic execution guidance
to further improve \tool's ability to detect promising patches during iterative repair.
We also aim to extend \tool to handle siblings not covered by
the existing test suite. To this end, we will investigate generating new tests to 
expose such siblings and leveraging these tests to guide their repair.

\bibliographystyle{IEEEtran}
\bibliography{ref}

\end{document}